\documentclass[twoside]{article}
\usepackage{fleqn,espcrc2,psfig}

\usepackage{graphicx}

\newcommand{\AmS}{{\protect\the\textfont2
  A\kern-.1667em\lower.5ex\hbox{M}\kern-.125emS}}

\begin{document}

\title{Collective Modes and Electronic Raman Scattering in the Cuprates}

\author{F. Venturini$^1$, U. Michelucci$^2$, T.P. Devereaux$^3$, and
A.P. Kampf$^2$\\
$^1$Walther-Meissner-Institut,
Walther-Meissner-Str. 8, 85748 Garching, Germany\\
$^2$Theoretische Physik III, Universit\"at
Augsburg, 86135 Augsburg, Germany\\
$^3$Department of Physics, University of
Waterloo, Waterloo, Canada, N2L 3G1\\
}       

\begin{abstract}
While the low frequency electronic Raman response in the
superconducting state of the cuprates can be largely understood in
terms of a d-wave energy gap, a long standing problem has been an
explanation for the spectra observed in $A_{1g}$ polarization
orientations. We present calculations which suggest that the
peak position of the observed $A_{1g}$ spectra is due to a collective spin
fluctuation mode. 

\end{abstract}

\maketitle

\section{INTRODUCTION}

In spite of the considerable efforts to explain the experimental Raman
spectra of cuprate superconductors, the $A_{1g}$ superconducting response
is not yet completely understood. It has been shown that the
theoretical description of the $A_{1g}$ Raman response 
was very sensitive to small changes in the Raman
vertex harmonic representations, yielding peak positions varying between
$\Delta$ and 2$\Delta$ \cite{tpdeinz}. However,
the data show peaks consistently slightly above 
$\Delta$ for both YBCO and BSCCO.

In this paper we
present calculations suggesting that the $A_{1g}$ peak position is largely
controlled by a collective spin fluctuation (SF)
mode near 41 meV, consistent 
with inelastic neutron scattering (INS) observations \cite{fong,mook}.
We show that the $A_{1g}$ response is strongly modified by the SF
term and is not sensitive to small changes in the Raman vertex.
The experimental peak position is well reproduced by
our model whereas the $B_{1g}$ and $B_{2g}$ response remain
essentially unaffected by the SF mode.

\section{MODEL CALCULATION}

The CuO$_2$ bilayer is modeled by a tight binding band structure with
a nearest ($t$) and a next nearest neighbor
hopping ($t'$) parameter and an inter-plane hopping given by \cite{bulut}
\begin{equation}
t_\perp({\bf k}) = 2 t_\perp \cos (k_z) [\cos(k_x)-\cos(k_y)]^2 .
\end{equation}
$k_z$ can be 0 or $\pi$, for bonding or anti-bonding bands
of the bilayer, respectively.

The spin susceptibility ($\chi_s$) is modeled by extending the weak
coupling form of a $d_{x^2-y^2}$ superconductor to include
antiferromagnetic spin fluctuations by an RPA form with an effective
interaction $\bar U$; i.e. $\chi_s=\chi_0/(1-\bar U\chi_0)$ where
$\chi_0$ is the simple bubble in the d-wave state.
This form of the spin susceptibility is motivated by the fact that it
contains a strong magnetic resonance peak at ${\bf q}={\bf Q}=(\pi,\pi,\pi)$
which was proposed \cite{bulut} to 
explain the INS resonance at energies near 41 meV in YBCO \cite{fong}
and BSCCO \cite{mook}.

\begin{figure}[hbt]
\centerline{%
\psfig{file=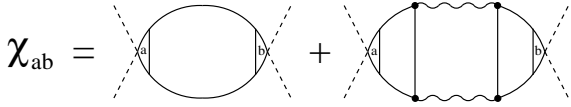,width=7.5cm}}
\caption{Feynman diagram considered for the particle-hole and SF 
contributions. Dashed, wiggly and solid lines
represent photon, SF and electronic propagators,
respectively. The solid circle marks the coupling $\bar U$ for the
electron-SF vertex.}
\label{total}
\end{figure}

The Raman response function in the superconducting state is evaluated
using Nambu Green's functions.
The spin fluctuations contribute to the Raman response via a 2-magnon
process as shown in Fig. \ref{total} \cite{kampf} where a schematic
representation of the Feynman diagrams of the SF and the bubble
contribution is plotted. 
For the electronic propagators we have used the bare BCS
Green's functions and a
d-wave superconducting gap 
$\Delta_{\bf k}=\Delta_0  [\cos (k_x) - \cos (k_y)]/2$.

The total Raman response is calculated in the gauge
invariant form which results from taking into account the long
wavelength fluctuations 
of the order parameter \cite{tpdeinz}. The total Raman susceptibility
is thus given by 
\begin{equation}
\chi_{tot}({\bf q=0},i\Omega)=\chi_{\gamma \gamma}({\bf 0},i\Omega)
- \frac{ \chi^2_{\gamma 1}({\bf 0},i\Omega)}{\chi_{11}({\bf 0},i\Omega)}
\label{scr1}
\end{equation}
where $\chi_{ab}({\bf q=0},i\Omega)$ is determined according to Fig.
\ref{total}. The analytical continuation to the real axis is performed
using Pad\'e approximants.

We have used several different forms for the
Raman vertex $\gamma$
which possess the correct transformation properties required
by symmetry.
Our calculations show that the SF term yields vanishingly small corrections to
the response in the $B_{1g}$ and $B_{2g}$ channels, but contributes
substantially to the $A_{1g}$ channel.
The shape of the total response in the $A_{1g}$ geometry is mainly
dependent on the value of the effective interaction $\bar U$. 
Variations of $\bar U$ change the relative magnitude of the two diagrams 
summed in Fig. \ref{total}, changing the position of the peak in 
$A_{1g}$ geometry.
Importantly, we find that the $A_{1g}$ response shows little dependence
on the form used for the vertex: $\cos(kx)+\cos(ky),
\cos(kx)\cos(ky)$, or the vertex calculated in an effective mass approximation.
These results can be explained by symmetry reasons 
given that the SF propagator is strongly peaked for ${\bf Q}$ momentum
transfers. 

\section{COMPARISON WITH DATA}

\begin{figure}[h]
\centerline{%
\psfig{file=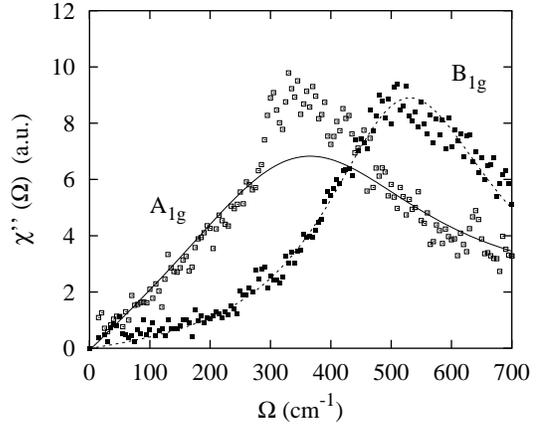,width=7cm}}
\vskip -1cm
\caption{Comparison of the $A_{1g}$ and $B_{1g}$ total response with Bi-2212
data taken from Ref. \cite{Barbara}.
The values of the parameters are $\Delta_0/t=0.25$, $t'/t=0.45$,
$t_\perp/t=0.1$, $\langle n \rangle = 0.85$, $\bar U/t=1.3$, $t=130$ meV.  
} 
\label{fig22}
\end{figure}

We compare the calculated Raman response with the
experimental spectra of an optimally doped Bi-2212 sample \cite{Barbara}
in Fig. \ref{fig22}.
Adding the SF
contribution leads to a shift of the peak position from near $\sim \Delta_0$ 
for $\bar U=0$ to higher
frequencies, allowing a better agreement with the experimental
relative positions of the peaks in $A_{1g}$ and $B_{1g}$ geometries. 
For the fit we have adjusted $t$ to achieve a good
agreement with the $B_{1g}$ channel, obtaining $t=130$ meV, and then
adjusted $\bar U$ to match both the $A_{1g}$ peak position as well as the peak
in the SF propagator to be consistent with the INS peak at 41 meV. 

From this work we conclude that including the SF contribution in the
Raman response solves the previously unexplained
sensitivity of the $A_{1g}$ response to small changes in the
Raman vertex.  Whereas the SF (two-magnon) contribution 
controls the $A_{1g}$ peak, the $B_{1g}$ and $B_{2g}$ scattering geometries are
essentially unaffected and determined by the bare bubble alone.

We would like to thank R. Hackl for numerous discussions.
One of the authors (F.V.) would like to thank the Gottlieb Daimler and 
Karl Benz Foundation for financial support.

\end{document}